\documentclass[%
 reprint,
 superscriptaddress,
 amsmath,amssymb,
floatfix,
]{revtex4-2}
\usepackage[utf8]{inputenc}

\usepackage{amsthm}
\newtheorem*{definition}{Definition}

\usepackage[normalem]{ulem} 

\usepackage{natbib}		
\setcitestyle{square,numbers}
\usepackage{color}
\usepackage[margin=1.0in]{geometry}

\usepackage[breaklinks=true,%
            colorlinks=true,%
            linkcolor=blue,%
            urlcolor=blue,%
            citecolor=blue]{hyperref}
            
\usepackage{amsmath}
\usepackage{amssymb}

\usepackage{bbold} 

\newcommand{\bra}[1]{\ensuremath{\left\langle#1\right|}}
\newcommand{\ket}[1]{\ensuremath{\left|#1\right\rangle}}

\newcommand{\ketbra}[2]{\ensuremath{\left|#1\right\rangle\left\langle#2\right|}}
\newcommand{\sandwich}[3]{\ensuremath{\left\langle#1\middle\vert#2\middle\vert#3\right\rangle}}

\usepackage{algpseudocode}
\usepackage{algorithm}

\usepackage{graphicx}
\usepackage{subfigure}
\graphicspath{ {figs/} } 

\usepackage{tikz}
\usetikzlibrary{quantikz}

\begin{document}

\title{Hybrid quantum-classical algorithms in
the noisy intermediate-scale quantum era and beyond}

\author{Adam Callison}
\email{a.callison@ucl.ac.uk}
\affiliation{Department of Physics and Astronomy, University College London, London, WC1E~6BT, UK}

\author{Nicholas Chancellor}
\email{nicholas.chancellor@gmail.com}
\affiliation{Department of Physics; Joint Quantum Centre (JQC) Durham-Newcastle, Durham 
University, South Road, Durham, DH1~3LE, UK}

\date{\today}

\begin{abstract}
Hybrid quantum-classical algorithms are central to much of the current research in quantum computing, particularly when considering the noisy intermediate-scale quantum (NISQ) era, with a number of experimental demonstrations having already been performed.
In this perspective, we discuss in a very broad sense what it means for an algorithm to be hybrid quantum-classical.
We first explore this concept very directly, by building a definition based on previous work in abstraction/representation theory, arguing that what makes an algorithm hybrid is not directly how it is run (or how many classical resources it consumes), but whether classical components are crucial to an underlying model of the computation. 
We then take a broader view of this question, reviewing a number of hybrid algorithms and discussing what makes them hybrid, as well as the history of how they emerged, and considerations related to hardware. 
This leads into a natural discussion of what the future holds for these algorithms. 
To answer this question, we turn to the use of specialized processors in classical computing.
The classical trend is not for new technology to completely replace the old, but to augment it.
We argue that the evolution of quantum computing is unlikely to be different: hybrid algorithms are likely here to stay well past the NISQ era and even into full fault-tolerance, with the quantum processors augmenting the already powerful classical processors which exist by performing specialized tasks.
\end{abstract}

\maketitle

\section{Introduction}
The first proposals for the creation of computing resources based on the principles of quantum mechanics emerged in the 1980s \citep{feynman1982simulating,benioff1980computer,deutsch1985quantum}.
This led quickly to the discovery, throughout the 1990s, of a number of quantum algorithms for toy problems designed to show a complexity theoretic separation between quantum and classical computation \citep{deutsch1992rapid,simon1997power,bernstein1997vazirani}.
In the mid-to-late 1990s, however, Peter Shor developed the polynomial-time quantum algorithm for factoring products of large prime numbers, \citep{shor1994algorithms, shor1999polynomial} now widely-regarded as the first \textit{useful} quantum algorithm that achieves a speedup over purely classical counterparts.

The explosion of interest in quantum computing that resulted from Shor's work is partly responsible for the rapid development thereafter of early experimental quantum hardware that could run small, noisy instances of quantum algorithms (see \citep{chuang1998experimental,brooke1999quantum,marx2000approaching,vandersypen2001experimental,gulde2003implementation,anwar2004implementation,negrevergne2006benchmarking,lu2007experimental,tame2007experimental,lu2009demonstrating,johnson2011quantum,devitt2016performing,arute2019quantum,zhong2020quantum} among others).
However, as these experimental platforms with severe limits on computation size and quality emerged, it quickly became apparent that for quantum computers to be useful in the near- and mid-term, the community needed to find algorithms tailored to this era of quantum computing (later termed the ``noisy intermediate-scale quantum" (NISQ) era \citep{preskill2018quantum}).
This includes not only algorithms run on gate model quantum devices, but also \textit{analog} optimization, simulation, and machine learning through quantum annealing \cite{kadowaki1998quantum}.

Perhaps the most influential development in the field of gate-model NISQ-suitable quantum algorithms is the framework of variational quantum algorithms (VQAs) that was introduced with the variational quantum eigensolver (VQE) \citep{peruzzo2014variational}.
These algorithms make use of short-depth parametrized quantum circuits, particularly suited to NISQ hardware, embedded in an otherwise classical variational loop, a structure that is manifestly \textit{hybrid quantum-classical}.
While some of the other quantum algorithms discovered previously are also hybrid, VQAs offered a clear demonstration on the potential computing power available by coupling quantum and classical resources together.
Thus, due to VQAs, hybrid quantum-classical algorithms have been an integral part of quantum algorithms research ever since, 
including through experimental demonstrations of applying NISQ devices to problems in quantum chemistry \citep{peruzzo2014variational,omalley2016scalable,kandala2019error}, machine learning \citep{havlicek2019supervised,larose2019variational}, and combinatorial optimization \citep{harrigan2021quantum}.

Separately from VQAs, the key advance which has allowed for hybrid techniques in \textit{analog} quantum computing is the notion of biased search \citep{perdomo-ortiz2011study,duan2013alternative,chancellor2017modernizing,grass2019quantum}, a family of techniques that allow the system to be called as a subroutine within a broader algorithm, aided by previously results \cite{chancellor2017modernizing,chancellor2016modernizing2}. 
These can be used in simple ways such as biasing toward a previously found good solution from a classical greedy search \citep{venturelli2019reverse}, or in more complicated ways such as in genetic algorithms \citep{king2019quantum}.

In this perspective, we take a broad view of hybrid quantum-classical algorithms, going beyond variational algorithms and hybrid quantum annealing.
We discuss what it means to be a hybrid algorithm, in the broadest sense of being necessarily a combination of quantum and classical computation.
We explore this question both in the literal sense of developing a definition and in the more abstract sense of where these algorithms sit within the space of quantum computing and more broadly computing in general. 

First, we propose that defining hybrid algorithms should be done in terms, not of the computation itself, but of the abstract model of the computation, and whether or not that model requires significant classical computation (computation which does not need a quantum hardware to be performed). 
We further examine how hybrid algorithms are developed and find that there are many paths, and that this development can be driven by any number of factors, including the capabilities and limitations of the hardware, the discovery of issues with ``pure" quantum algorithms, and the identification of bottlenecks in current classical methods.

Furthermore, we consider the long and successful history of specialized processors being used within classical computing and note that, while quantum computers can be very powerful and greatly enhance computational capabilities, it is likely that they will play a role similar to many of these other accelerators. 
Finally, looking to the future, we see no clear advantage in algorithms being ``purely'' quantum, and therefore argue that hybrid algorithms are likely here to stay, rather than just an artifact of the NISQ era.

We start in section \ref{sec:what_it_means} by discussing how a useful definition can be constructed, and how to make the distinction between pure and hybrid quantum algorithms meaningful.
Within this discussion, we reference two of the most iconic quantum algorithms: Shor's algorithm, which we argue is in fact hybrid, and Grover's algorithm, which we argue is not.
In section \ref{sec:alg_examples}, we discuss examples of hybrid quantum-classical algorithms and how they fit into the definition. 
Next, in section \ref{sec:hardware} we examine the considerations that must be made when thinking of how hybrid algorithms will be run on real hardware. 
We then, in section \ref{sec:specialized_accelerators}, examine the bigger picture; in particular, how classical computation makes use of heterogeneous hardware arrangements as specialized accelerators and argue that quantum processors are likely to follow as similar path.
Finally, in section \ref{sec:future}, we conclude by discussing the future of hybrid algorithms. 

\section{What it means to be hybrid \label{sec:what_it_means}}

To begin a meaningful discussion of hybrid quantum-classical algorithms, it is first necessary to establish an understanding of what it means for a quantum algorithm to be considered a hybrid algorithm and what makes hybrid algorithms different from other quantum algorithms. 
For this task, it is essential to develop a working definition of the term ``hybrid'' in this context. 
An agreed upon, meaningful and useful definition can help facilitate communication and allow a useful grouping of similar algorithms. 
It is tempting to argue that simple techniques such as ``repeat until success'' play an important role in almost all quantum algorithms, and therefore almost all quantum algorithms should be considered hybrid. 
While potentially a valid viewpoint, this approach does not lead to a useful definition: if everything is hybrid, then the term ``hybrid quantum algorithm'' has little meaning. 

Rather than considering to be hybrid any quantum algorithm for which the implementation is supported by classical computing, it seems that the salient feature connecting hybrid quantum-classical algorithms is that the use of classical processing is fundamentally inseparable from the computational model.
While we will discuss later in this section what exactly this term means, for now it suffices to think of this as the basic theoretical description of how a system solves problems.
As a starting point, therefore, we propose the following definition of a hybrid quantum-classical algorithm:
\begin{definition}
An algorithm that requires non-trivial amounts of both quantum and classical computational resources to run, and which cannot be sensibly described, even abstractly, without reference to the classical computation.
\end{definition}
While the term ``non-trivial" in this context is open to interpretation, a sensible interpretation is that the classical computation should go beyond simple repetition and conditioning on measurement outcomes.

\begin{figure}[ht]
\includegraphics[width=0.5\columnwidth]{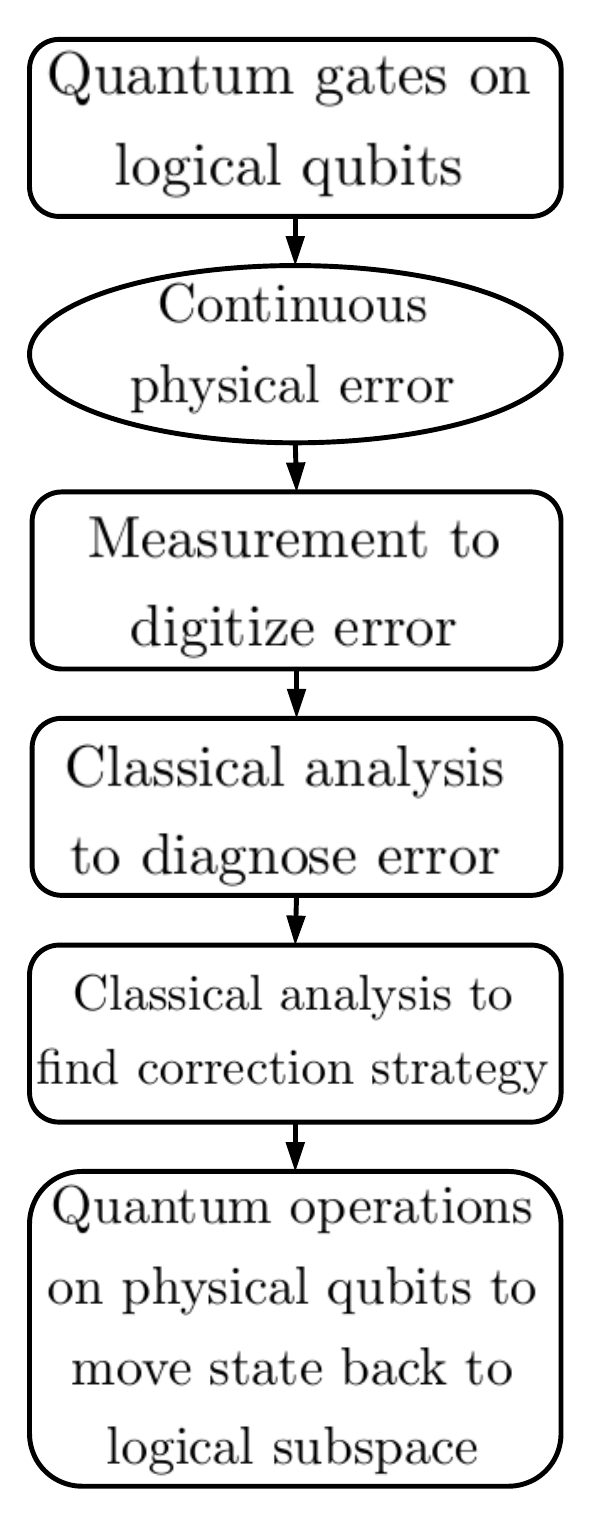}
\caption{\label{fig:qec} 
High-level schematic of typical forms of QEC in which some classical processing is used to protect a quantum algorithm from errors.
}
\end{figure}

Quantum error correction \cite{roffe2019quantum} (QEC) provides a simple example of classical computers supporting quantum algorithms, but in which the classical computation is not fundamentally tied into the computational model. 
In devices with QEC, the quantum computation occurs only on a small part of the device's Hilbert space, known as the \textit{logical subspace}.
The logical subspace is chosen such that almost all physical errors will take the system out of this subspace; that is, errors within the subspace are unlikely.
Well-chosen measurements than can be performed to detect whether the state has left the logical subspace without destroying the quantum information (more correctly, the error will lead to a superposition over the logical subspace and the error subspaces, but the measurement will force the state to collapse onto one of the subspace, a mechanism referred to as \textit{error digitization}).
Classical analysis on these measurement outcomes can then be used to diagnose the error and to determine the quantum operations required to rotate the state back into the logical subspace, in a mechanism known as \textit{decoding}.
The high-level structure of typical QEC strategies is shown in Fig. \ref{fig:qec}, and a thorough review can be found in \citep{roffe2019quantum}.

Decoding of quantum error correction can be very intensive in terms of classical computing resources \cite{breuckmann2017local}; however, the fundamental model of (noise-free) gate-model quantum computing does not depend on \emph{how} the errors are corrected. 
Similarly, the classical strategy for decoding does not depend on exactly what quantum algorithm is running, only on the device architecture.
We would therefore argue that the fact that classical computing resources are required to protect a fault tolerant quantum algorithm does not in itself make the algorithm hybrid. 
A fault tolerant algorithm can be hybrid in the sense that the quantum component of the algorithm can be run in a fault-tolerant way, but the fact that the algorithm relies on QEC does not in itself make it hybrid.

To approach the question of when a quantum computation is hybrid, it is useful to first look at work which has already been done in understanding when a physical system computes. 
The Abstraction-Representation theory \cite{horsman2018abstraction} of computation was motivated by exactly this question \cite{horsman2014when}.
The key distinction which abstraction representation theory draws between a physical system which is computing and one which is not is that a system undergoing computation has information encoded in it and a corresponding abstract model for the computation it is doing. 
In other words, computation is not about what a system is doing so much as what it is \emph{representing}. 

By abstraction-representation theory, a physical system needs to be paired with a (possibly imperfect) abstract model of the computation it is doing. 
For example, a gate-model quantum computer can be abstractly represented by a quantum circuit (possibly including random elements from noise), a coherent quantum annealer could be represented by the Schr{\"o}dinger equation, and a dissipative quantum annealer by a master equation. 

To understand whether a computation is hybrid or not, it is necessary to think in terms of what the role classical computation plays within this model. 
Returning to the example of QEC, the model for fault-tolerant quantum computing is already a (noise free) quantum circuit, and the classical computation is just performing a supporting role in ensuring that the model is accurate enough. 
For contrast, consider a variational algorithm; in this case the classical computation is directly involved within the computational model, it gives the updates for the gate parameters and fundamentally cannot be removed from the computational model.

To demonstrate the differences between pure quantum algorithms and hybrid quantum-classical algorithms, we now turn to two widely-known algorithms: Grover's algorithm for searching unsorted databases, and Shor's algorithm for integer factorization.
In what follows, we argue that Grover's algorithm should not be considered meaningfully hybrid, while Shor's algorithm should. 
We discuss other example of hybrid algorithms in more detail in section \ref{sec:alg_examples}.

\subsection{Grover's algorithm}
Grover's algorithm \citep{grover1996fast} is a well-known quantum algorithm for solving the `unstructured search' problem. 
For the purpose of this section we limit ourselves to ``pure'' implementations of Grover's algorithm, where the entire problem is phrased as an unstructured search; we discuss more sophisticated algorithms based on the same amplitude amplification principle as Grover's algorithm in section \ref{subsec:QAE}.
Very simply, this is the problem of finding any of $M$ marked labels among a set of $N \geq M$ of labels, given knowledge of the numbers $M$ and $N$ access to an oracle that is able to check whether a particular label $j$ is marked.
Many computational problems can be phrased as unstructured search, such as database search and some forms of constraint-satisfaction.
Without access to additional structure among the labels, the only option for solving unstructured search classically is to repeatedly select labels, either randomly or in a particular order, which has a time complexity that scales as $O\left(\frac{N}{M}\right)$.  

The labels $j$ are mapped to computational basis states $\ket{j}$ of a Hilbert space of size $N$, and we assume \textit{coherent} access to the oracle; that is, it is possible to apply a unitary operator $O_\mathrm{check}$ defined by
\begin{eqnarray}
    O_\mathrm{check}\ket{j} = \left\{
    \begin{array}{lr}
        -\ket{j} & \mathrm{for\ marked\ }j\\
        \ket{j} & \mathrm{for\ unmarked\ }j.
    \end{array}\right.
\end{eqnarray}
It can be shown that by initialising the quantum computer in the uniform superposition state $\ket{s}=\frac{1}{\sqrt{N}}\sum_{j=1}^{N}\ket{j}$ and applying $r$ iterations of the Grover operator
\begin{eqnarray}
    Q_\mathrm{Grover} &=& \left(2\ketbra{s}{s} - \mathbb{I}\right)O_\mathrm{check},\label{eq:grover}
\end{eqnarray}
a state is produced that can be measured in the computational basis to return a marked state with probability
\begin{eqnarray}
    P_\mathrm{marked}(r) &=& \sin^2\left((2r+1)\theta \right)
\end{eqnarray}
where $\theta=\sqrt{\frac{M}{N}}$.
Thus, by choosing
\begin{eqnarray}
    r=\mathrm{Round}\left(\frac{\pi}{4}\sqrt{\frac{N}{M}}-\frac{1}{2}\right)\in O \left( \sqrt{\frac{N}{M}} \right),
\end{eqnarray}
a marked state can be found with a probability that quickly approaches unity for $M\ll N$.
This represents a quadratic speedup over what is possible classically.

While single run of the algorithm does not guarantee a marked state with unit probability, the output can be checked via the oracle and the algorithm repeated a small number of times until a marked state is confirmed, without affecting the quadratic speedup.
It would not be useful to consider a quantum algorithm with this trivial level of classical post-processing to be hybrid, as it is difficult to imagine any useful quantum algorithms which did not contain at least these elements of classical control.

\subsection{Shor's algorithm}
\label{ssec:shor}
Shor's polynomial-time algorithm \citep{shor1999polynomial} for factoring products $N=p_1p_2$ of two large prime numbers $p_1$ and $p_2$, a feat thought to be impossible classically, is regarded as the first useful quantum algorithm to have been invented.
Given the risk a practical implementation of Shor's algorithm would have on much of modern computer security, it is perhaps the most high-profile quantum algorithm, and is seen as a quintessential example of the power of quantum computing.

\begin{figure}
\begin{centering}
\resizebox{0.99\columnwidth}{!}{
\begin{quantikz}
 \lstick{\ket{0}} & \gate{H} & \ctrl{5} & \qw & \qw & \ \ldots\ \qw & \qw & \gate[5, nwires=4]{\mathrm{QFT}^{-1}} & \meter{} \\
 \lstick{\ket{0}} & \gate{H} & \qw & \ctrl{4} & \qw & \ \ldots\ \qw & \qw & & \meter{}\\
 \lstick{\ket{0}} & \gate{H} & \qw & \qw & \ctrl{3} & \ \ldots\ \qw & \qw & & \meter{}\\
 \lstick{\ \vdots\ } & \ \vdots\  & & & & & & & \ \vdots\ \\
 \lstick{\ket{0}} & \gate{H} & \qw & \qw & \qw & \ \ldots\ \qw & \ctrl{1} & & \meter{}\\
 \lstick{\ket{\psi}} & \qwbundle
{n} & \gate{U^{2^0}} & \gate{U^{2^1}} & \gate{U^{2^2}} & \ \ldots\ \qw & \gate{U^{2^{(m-1)}}} & \qw & \qw \\
\end{quantikz}
}
\caption{\label{fig:phaseestcirc} 
Phase estimation circuit.
}
\end{centering}
\end{figure}
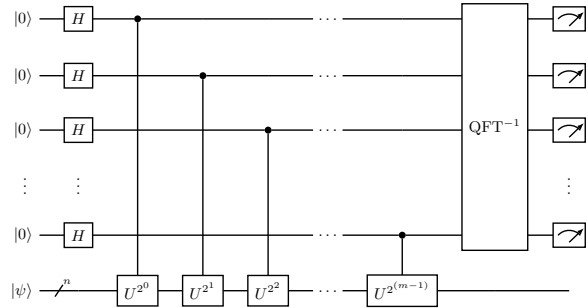

However, Shor's algorithm is far from being a \textit{purely} quantum algorithm, and should certainly be considered a hybrid algorithm.
The algorithm relies on a polynomial-time reduction of the factoring problem to the problem of finding the order $r$ of a periodic function, and is described in Algorithm \ref{alg:shor}.
It can be seen that almost all of the steps are entirely classical, including the calculation of greatest common divisors via the classical Euclidean algorithm in steps \ref{algstep:gcd1} and \ref{algstep:gcd2} and the classical continued-fraction expansion in step \ref{algstep:confrac}.
The only part of the algorithm that is quantum is the call to the quantum phase estimation circuit (see Fig. \ref{fig:phaseestcirc}) in step \ref{algstep:runquantumcircuit}, which forms the core of the order-finding subroutine.

\begin{algorithm}[H]
\caption{Structure of Shor's algorithm\label{alg:shor}}
\begin{algorithmic}[1]
\State Randomly select $a$ from $1<a<N$ \label{algstep:randomselection}
\State Compute $K=\mathrm{gcd}(a,N)$ \textbf{classically}, via Euclidean algorithm\label{algstep:gcd1}
\State If $K\neq1$, $K$ is a factor of $N$, so return $a$ and exit. Otherwise, continue to step \ref{algstep:implementunitary}.
\State Implement the $a$-dependent unitary $U_a$ that acts as $U_a\ket{x}_n=\ket{a^x\,\mathrm{mod}\,N}$.\label{algstep:implementunitary}
\State Perform \textbf{quantum} phase estimation (QPE), via a circuit like the one shown in Fig. \ref{fig:phaseestcirc}, with a $q$-qubit control register where $N^2<2^q<2N^2$, on $U_{a}$ with the initial state of the target register set to $\ket{1}$ (the binary representation of 1), to produce an integer $y$. \label{algstep:runquantumcircuit}
\State Use \textbf{classical} continued-fraction expansion to produce an irreducible fraction approximation $\frac{d}{r}$ of $\frac{y}{2^{q}}$ for which $r<N$ and the approximation error is small ($<\frac{1}{2^{q+1}}$). \label{algstep:confrac}
\State If $r$ is odd or it is true that $a^{\frac{r}{2}}=-1\,\mathrm{mod}\,N$, go back to step \ref{algstep:randomselection}. Otherwise $r$ is highly likely to be the period of the modular exponentiation with base $a$, $f(x)=a^x\,\mathrm{mod}\,N$, so continue to step \ref{algstep:checkifperiod}.
\State If $r$ \textit{is} the period of the modular exponentiation (which can be checked \textbf{classically}), continue to step \ref{algstep:gcd2}.
Otherwise, go back to step \ref{algstep:confrac}, and find a different irreducible fraction with a different $r$.
If out of candidates, go back to step \ref{algstep:randomselection}.\label{algstep:checkifperiod}
\State Compute $\mathrm{gcd}(a^{\frac{r}{2}}-1,N)$ and $\mathrm{gcd}(a^{\frac{r}{2}}+1,N)$ \textbf{classically}, via the Euclidean algorithm. 
These are now guaranteed to be non-trivial factors of $N$, so return these and exit.\label{algstep:gcd2}
\end{algorithmic}
\end{algorithm}

While it is absolutely true that the quantum part of Shor's algorithm provides the speedup over the best known purely classical methods, a significant (but polynomial) amount of classical processing is done before and after the call to the quantum subroutine in step \ref{algstep:runquantumcircuit} in order to transform the problem into a specific form that the quantum processor can handle efficiently. 
This classical processing includes calls to identifiable classical algorithms such as the Euclidean algorithm, 
In other words, what is being calculated by the quantum computer is not directly useful without the surrounding classical steps. 

Having discussed two of the possibly most iconic quantum algorithms, in the next section we get into a more detailed discussion about other important examples of hybrid algorithms.
This is not intended to be an exhaustive list, but to start a discussion on what it means to be hybrid under different contexts, and to provide examples that can help extend these to other algorithms.

\section{Algorithm examples \label{sec:alg_examples}}

\subsection{Quantum Annealing and Continuous-Time Quantum Computing}
\label{ssec:annealing}
Quantum annealing (QA) consists of using continuous-time dynamics of a quantum system to solve an optimization problem by searching the solution space through quantum fluctuations. 
QA as we know it today was proposed in the late 1990s by Kadawaki and Nishimori \citep{kadowaki1998quantum} (although a similar suggestion for chemistry related problems had been made earlier \citep{finnila1994quantum}). 
After the development of the basic principles of QA, Farhi et al. introduced the specific case of adiabatic quantum computing (AQC) \citep{farhi2000adiabatic} in which the adiabatic theorem of quantum mechanics guarantees a solution if the algorithm is run long enough.
Roland and Cerf than showed that AQC can attain the same speedup as the gate model Grover's algorithm on unstructured search \citep{roland2002quantum}, while around the same time Santoro et al. further developed the theory of QA and showed that it can in some cases outperform classical annealing algorithms \citep{santoro2002theory}.
For a complete review of AQC, see \citep{albash2018adiabatic}, and for a more application-focused review of QA, see \citep{yarkoni2021quantum}. 
For a forward looking perspective on techniques which operate outside of the adiabatic limit see \citep{crosson2021prospects}.

Since it's inception, QA has grown to encompass a variety of protocols, but the typical structure involves implementing a time-dependent transverse Ising model Hamiltonian
\begin{eqnarray}
    H(t) &=& -A\left(\frac{t}{T}\right)\sum_{j=1}^{n}X_j + B\left(\frac{t}{T}\right)H_P\label{eq:qa_ham}
\end{eqnarray}
where $T$ is the total annealing time, $A$ and $B$ are monotonic control functions such $A(0)\gg B(0)$ and $A(1)\ll B(1)$, and the `problem Hamiltonian' $H_P$ is of the form
\begin{eqnarray}
    H_P & = & -\sum_{j=1}^{n-1}\sum_{k=j+1}^{n}J_{jk} Z_j Z_k - \sum_{j=1}^n h_j Z_j
\end{eqnarray}
where the couplings $J_{jk}$ and fields $h_j$ define the problem of interest such that lower-energy eigenstates of $H_P$ correspond to better quality solutions.
In this way, QA is qualitatively similar to its classical analog \textit{simulated} annealing, but with the physical quantum fluctuations/tunneling induced by the transverse field `driver' term $-\sum_{j=1}^{n}X_j$ playing the role of a simulated temperature.
QA protocols typically start the system in the ground state of the transverse-field driver term, and aim to set the controls functions $A$ and $B$ such that the system is driven toward low energy states of the problem Hamiltonian.

\subsubsection{Hybrid forms of quantum annealing \label{subsubsec:hybrid_anneal}}

Much of the work on hybrid algorithms in continuous-time quantum computing has focused on protocols which allow an educated guess at the solution to be incorporated into the protocol. 

The idea of incorporating a heuristic guess was first proposed by Perdomo-Ortiz et al. \citep{perdomo-ortiz2011study}, in a scheme which has come to be known as (one variant of) \textit{reverse} annealing.
In this scheme, the standard QA Hamiltonian of Eq. \eqref{eq:qa_ham} is replaced with
\begin{eqnarray}
    &H_{\mathrm{rev},1}(t) = \left[1-s\left(\frac{t}{T}\right)\right]H_\mathrm{init} +& \nonumber \\
    &\mathrm{hat}\left[s\left(\frac{t}{T}\right)\right]\left(-\sum_{j=1}^{n}X_j\right) + s\left(\frac{t}{T}\right)H_P, & \label{eq:revanneal1}
\end{eqnarray}
where $s$ is a \textit{schedule} function that starts at $s(0)=0$, ends at $s(1)=1$ and monotonically increases in between.
The $\mathrm{hat}$ function pre-multiplying the transverse field driver is a function that begins and ends at zero, but is positive in between, such as $\mathrm{hat}(s)=\sin(\pi s)$; that is, the transverse field driver is turned on and then off during the anneal.
The completely new ingredient is the \textit{initial} Hamiltonian 
\begin{eqnarray}
    H_\mathrm{init} &\equiv& -\sum_{j=1}^{n}g_j Z_j,
\end{eqnarray}
where $g_j$ is a sign such that $g_j=1$ ($g_j=-1$) if the $j$th bit of the heuristic guess is $0$ ($1$).
The ground-state of the initial Hamiltonian $H_\mathrm{init}$ is then the guess state, and the annealing protocol aims to transfer a non-trivial amount of the amplitude into the true ground-state of the problem Hamiltonian $H_P$, with the transverse field turned up and then down in order to assist this transfer via quantum fluctuations.
It is worth noting that, since the aim of a reverse annealing run is to improve upon a a previously known reasonable solution, it is possible to embed reverse annealing in a classical loop, whereby the output state measured one run is then incorporated as the initial state (with the appropriate initial Hamiltonian) for the next run.
This hybrid structure was suggested in \citep{perdomo-ortiz2011study}, and schemes of this sort have come to be known as \textit{iterated} reverse annealing.

Other protocols have been proposed that have also come to be known as reverse annealing.
For example, Chancellor \citep{chancellor2017modernizing} and Yamashiro et. al \citep{yamashiro2019dynamics} propose using the standard QA Hamiltonian of Eq. \eqref{eq:qa_ham}, but simply initializing to the guess state and using a schedule that begins entirely in the problem Hamiltonian $H_P$ with $A(0)=0,B(0)=1$, anneals to some intermediate values $0 < A(t_\mathrm{intermediate}/T), B(t_\mathrm{intermediate}/T) < 1$, and then back to $A(1)=0,B(1)=1$.
These protocols often include pausing the schedule for some time at the intermediate values \citep{marshall2019power}.
In \citep{yamashiro2019dynamics}, this two-term form of reverse annealing is embedded in a classical loop and is also called \textit{iterated} reverse annealing (in fact, this is where the term  was coined).

A third form of reverse annealing, proposed by Ohkuwa et. al \citep{ohkuwa2018reverse}, uses a Hamiltonian of the form 
\begin{eqnarray}
    &H_{\mathrm{rev},2}(t) = \left[1-s\left(\frac{t}{T}\right)\right]\left[1-\lambda\left(\frac{t}{T}\right)\right]H_\mathrm{init} +& \nonumber \\
    &\left[1-s\left(\frac{t}{T}\right)\right]\lambda\left(\frac{t}{T}\right)\left(-\sum_{j=1}^{n}X_j\right) + s\left(\frac{t}{T}\right)H_P, & \label{eq:revanneal2}
\end{eqnarray}
where $s(0),\lambda(0) = 0$, $s(1),\lambda(1) = 1$, and both are monotonically increasing between.
Recognising that the system begins in the instantaneous ground-state at $t=0$ and is targeting the instantaneous ground-state at $t=T$, this protocol can in principle be run adiabatically, and has come to be known as \textit{adiabatic} reverse annealing.
This Hamiltonian in Eq. \eqref{eq:revanneal2} is quite similar to that described by Eq. \eqref{eq:revanneal1} (in principle, the Eq. \eqref{eq:revanneal1} protocol could also be run adiabatically), but the additional freedom that arises from having two schedule functions, $s$ and $\lambda$, offers additional flexibility for the annealing path.

An independent proposal that incorporates a heuristic guess by using a biased driver Hamiltonian, instead of by reverse annealing, was given by Duan et al. \citep{duan2013alternative} (later studies of this protocol can be found in \cite{grass2019quantum} and to a lesser extent in \cite{callison2021energetic}).
An example of such a protocol is to modify the transverse-field driver term to include a bias towards a particular guess state; that is, 
\begin{eqnarray}
    -\sum_{j=1}^{n}X_j &\rightarrow& -\sum_{j=1}^{n}\left(X_j + g_jb_jZ_j \right)
\end{eqnarray}   
where $b_j$ is a positive number representing the strength of the bias and $g_j=1$ ($g_j=-1$) is the $j$th bit of the guess is $0$ ($1$).
The initial state should also be biased towards the guess state, such that it remains the ground-state of the biased driver.

Gra\ss \hspace{0pt} \citep{grass2019quantum} showed numerically that biasing both toward and away from solution candidates can improve performance of annealing protocols, and has also experimentally demonstrated that biased-driver QA can work in practice \citep{grass2022quantum} (although technical limitations meant the bias had to be included as part of the problem Hamiltonian instead).
Callison et al. \citep{callison2021energetic} demonstrated that traditionally formulated QA is subject to an energy redistribution mechanism if certain conditions are met, and the conditions are met in the biased-driver protocols proposed of \cite{duan2013alternative,grass2019quantum} (but not in the reverse annealing protocols).

Since both reverse QA and biased-driver QA fundamentally incorporate classical information, in the form of previous knowledge of the solution, into the protocol itself, any algorithm which makes use of them is necessarily hybrid.
Many of the earlier papers focused on how closed system quantum annealing could incorporate prior knowledge, but later work explored how these ideas can be used in more sophisticated algorithms that use QA as a subroutine.
An influential example is the work by Chancellor \citep{chancellor2017modernizing}, in which it was proposed how a known experimental protocol (see: \cite{lanting2014entanglement}) on a dissipative quantum annealer could be used within analogues of classical population annealing \citep{matcha2010population,wang2015population} and parallel tempering \citep{swendsen1986parallel,earl2005parallel} algorithms. 
Furthermore, Chancellor \citep{chancellor2016modernizing2} proposed a general formalism to understand annealing protocols that accept initial conditions, such as reverse anneals.
Within a week of the first appearance of this paper on the ar$\chi$iv pre-print repository, Neven independently proposed a QA-based parallel tempering algorithm in the AQC 2016 opening remarks \citep{neven2016aqc}. 
After this, D-Wave Systems Inc.~made the necessary controls to perform a reverse annealing protocol (and, in fact, coined the term ``reverse annealing" \citep{dwave2019reverse}) using a similar method to the proposal in \citep{chancellor2017modernizing}.
The D-Wave implementation of reverse annealing most closely resembles the two-term proposals of \citep{chancellor2017modernizing,yamashiro2019dynamics}, except that it operates in the dissipative regime.  

The capability to perform reverse annealing has lead to numerous experimental studies. 
Venturelli and Kondratyev \citep{venturelli2019reverse} demonstrated that incorporating the result of a greedy search can improve the performance of a quantum annealer on a portfolio optimization problem. 
Ottaviani and Amendola \citep{ottaviani2018low} as well as Golden and O'Malley \citep{golden2021reverse} demonstrated that reverse annealing can aid in matrix factorization. 
Chancellor \citep{chancellor2020fluctuation} demonstrated how these techniques can be combined with other advanced controls to trade off between optimality and flexibility within solutions. 
Chancellor and Kendon \citep{chancellor2021experimental} demonstrated that the underlying principle behind using reverse annealing for local search was correct and examined the effect of noise levels.

\begin{figure*}[ht]
\includegraphics[width=1.5\columnwidth]{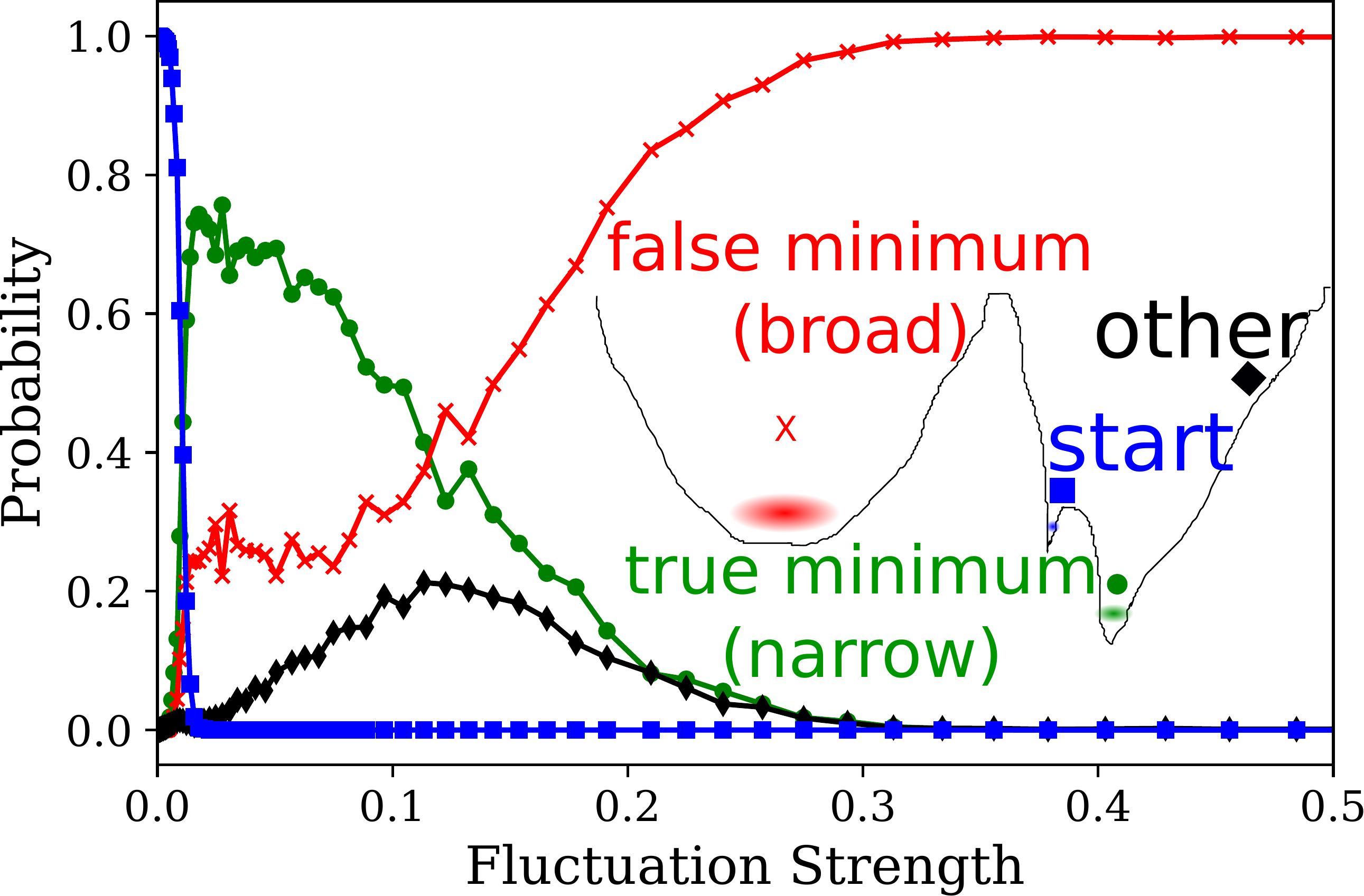}
\caption{\label{fig:reverse_annealing_demonstration} 
Experimental demonstration of dissipative reverse annealing with a cartoon representation of an engineered energy landscape in the inset. 
Biasing toward a true solution leads to the problem being solved with a high probability, while traditional forward annealing (approximated in the limit of large fluctuation strength) does not solve the problem. 
Data plot reproduced from \citep{chancellor2021experimental}, cartoon illustration is original. The colors (shades of gray in print) and symbols in the cartoon reproduce those in the figure. The cross symbol indicates probability to be found in the false minimum, while the circles represent the probability to be found in the true minimum. The squares indicate probability to be found in the original starting state, and diamonds indicate probability to be found in all states not previously mentioned.
}
\end{figure*}

Data from an experimental demonstration of using reverse annealing for local searching, first presented in \citep{chancellor2021experimental}, are shown in Fig.~\ref{fig:reverse_annealing_demonstration}. 
These experiments were performed by constructing an Ising system with known pathological behaviour for QA.
This was achieved by creating a ``broad'' false minimum where strong quantum fluctuations were allowed.
The system was than initialized near (in the sense of number of bitflips) a much ``narrower'' true minimum (see the cartoon inset of Fig.~\ref{fig:reverse_annealing_demonstration}).
The results demonstrate that searching the solution locally by starting from a `good guess' and accessing only moderate fluctuations through reverse annealing) allowed the broad false minimum to be avoided, thereby validating reverse annealing as a tool for search solution spaces locally, a key ingredient for hybrid QA algorithms.

A clear demonstration of a hybrid quantum-classical algorithm employing reverse annealing as a quantum (although not fully-coherent) subroutine was provided by King et al. in \citep{king2019quantum}.
In that work, reverse annealing is used as the mutation operator in an otherwise classical genetic algorithm.
The algorithm is shown in Algorithm \ref{alg:qaga}.

\begin{algorithm}[H]
\caption{Quantum-Assisted genetic algorithm of \citep{king2019quantum}\label{alg:qaga}}
\begin{algorithmic}[1]
\State \textbf{classically} initialize the \texttt{population} as $N$ random states
\State FOR \texttt{generation} IN \texttt{generations}:
\State \hspace{7pt} FOR \texttt{individual} IN \texttt{population}:
\State \hspace{14pt} with probability \texttt{mutation\_rate}, use \textbf{quantum} reverse annealing, starting from \texttt{individual} to create \texttt{mutated\_individual}
\State \hspace{14pt} add \texttt{mutated\_individual} to \texttt{population}
\State \hspace{7pt} END FOR
\State \hspace{7pt} randomly match pairs of individuals to create list \texttt{pairs} (of size $\texttt{recombination\_rate}\times|\texttt{population}|$)
\State \hspace{7pt} FOR \texttt{pair} IN \texttt{pairs}:
\State \hspace{14pt} \textbf{classically} combine \texttt{pair} to create \texttt{offspring}
\State \hspace{14pt} add \texttt{offspring} to \texttt{population}
\State \hspace{7pt} END FOR
\State \hspace{7pt} exit loop if a stopping condition is met
\State \hspace{7pt} discard old individuals from \texttt{population} to maintain desired size $N$
\State END FOR
\end{algorithmic}
\end{algorithm}

Perhaps most surprisingly, reverse annealing has aided not just optimization, but also quantum simulation using quantum annealers; in particular, demonstrations that quantum annealers can be used to simulate geometrically frustrated spin systems. 
The quantum simulations by King et al. \citep{king2018observation,king2021scaling} are only feasible through use of the reverse annealing feature to initialize the system in a previously found configuration. 
This is of particular interest since the work in \cite{king2021scaling} shows potential evidence of a scaling advantage over classical techniques based on path-integral quantum Monte Carlo.

There has also been significant numerical and theoretical work on understanding reverse annealing.
Ohkuwa et al. \cite{ohkuwa2018reverse} showed that reverse annealing techniques could mitigate or even remove a first order phase transition in highly symmetric problems, though the analysis assumes a coherent rather than dissipative setting.
In a similar vein, Yamashiro et al. \citep{yamashiro2019dynamics} showed that a more coherent version of reverse annealing works better than dissipative reverse annealing for solving highly symmetric problems, and discussed how protocols similar to those proposed here could be experimentally implemented with an existing feature on D-Wave devices.
Furthermore, in a pair of papers considering the $p$-spin model as the problem Hamiltonian, Passarelli et. al show that open quantum system effects can enhance the performance of \textit{iterated} reverse annealing \citep{passarelli2020reverse} (using a Hamiltonian of the form of Eq. \eqref{eq:qa_ham} with a nonmonotonic schedule), but could be detrimental in the case of \textit{adiabatic} quantum annealing \citep{passarelli2022standard} (using a Hamiltonian of the form of Eq. \eqref{eq:revanneal2}).

Overall, reverse annealing has been identified as a key technique to eventually realising a quantum advantage through annealing techniques, especially in settings where coherence is low.
In fact, in a recent forward-looking perspective, Crosson and Lidar \citep{crosson2021prospects} only rank QA in a moderately decoherent setting as promising if reverse annealing techniques are used.

Moving away from strategies that incorporate prior knowledge into the anneal, there are a number of other approaches to hybridizing QA, many of which involve optimizing QA schedules to increase success probability without requiring the runtime to grow toward the adiabatic limit, and fall under the umbrella of \textit{shortcuts to adiabaticity} \citep{guery-odelin2019shortcuts}.
For example, in \citep{hartmann2019rapid}, a parameterized counter-diabatic term is introduced to the Hamiltonian to mitigiate harmful aspects of the diabatic evolution, and its parameters are variationally-optimized in a classical loop, similar to variational algorithms in the gate model (see subsection \ref{ssec:variational}).
Related work is presented in \citep{passarelli2020counterdiabatic,passarelli2022optimal} 

An alternative approach to hybridizing QA is to use classical machine-learning techniques to design annealing schedules for particular problems or problem types.
For example, Chen et. al \citep{chen2022optimizing} show that a combination of tree-search algorithms and neural networks can be used to learn effective annealing schedules that appear to have the same effect as deliberately including counter-diabatic driving terms and which can be transferred between instances.
A genetic algorithm is used for similar purposes by Hegde et. al in \citep{hedge2022genetic}.

\subsection{Variational algorithms}
\label{ssec:variational}
\subsubsection{Variational Quantum Eigensolver} 
\label{sssec:vqe}

The NISQ era of quantum computing is characterized by devices with limited spatial (e.g number of qubits) and temporal (e.g achievable coherence time) resources.
As such, near-term quantum algorithm design must focus on extracting as much performance as possible out of these limited resources.
Hybrid algorithms offer a natural way to do this, by performing much of the computational effort on classical hardware and calling the quantum processor only for small subproblems to which it is particularly well-suited.
A well-known and illustrative example is the Variational Quantum Eigensolver (VQE) \citep{peruzzo2014variational}, a class of algorithms intended for NISQ devices that combines small quantum circuits and classical optimization techniques to approximate eigenstates of a Hamiltonian $\mathcal{H}$, typically ground-states, and associated eigenenergies.
Comprehensive overviews of VQEs and related variational algorithms are available elsewhere \citep{cerezo2021variational,endo2021hybrid}, but we give a brief description here to illustrate a key example of a hybrid algorithm.

\begin{figure}[ht]
\includegraphics[width=0.99\columnwidth]{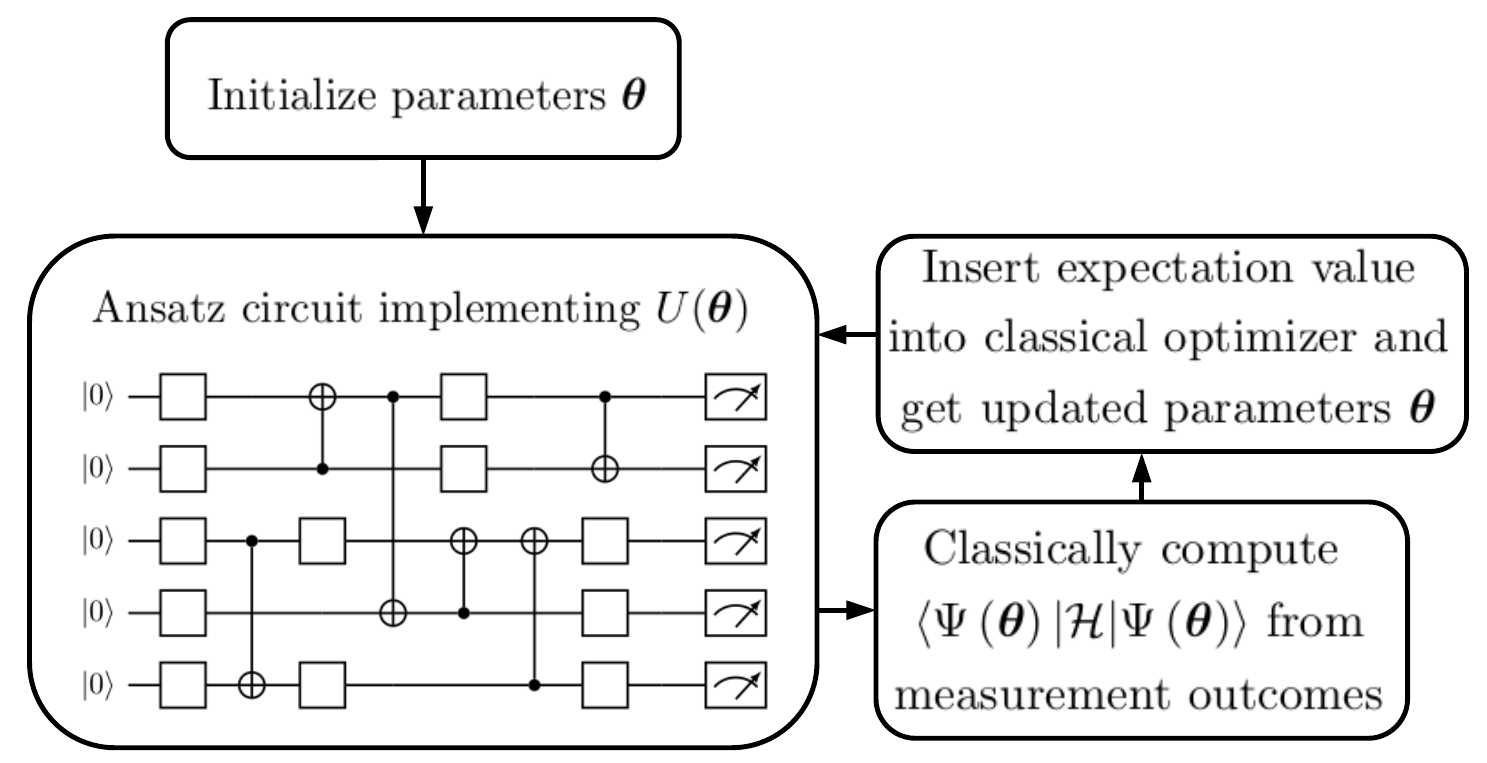}
\caption{\label{fig:vqe_diagram} 
Schematic showing a high-level overview of the structure of VQE algorithms.
}
\end{figure}

In VQE, a quantum processor is used to prepare a state 
\begin{eqnarray}
    \ket{\Psi(\boldsymbol{\theta})} &=& U(\boldsymbol{\theta})\ket{\Psi_\mathrm{init}}
\end{eqnarray}
where $\ket{\Psi_\mathrm{init}}$ is an easily-prepared initial state.
The unitary operator $U(\boldsymbol{\theta})$ is the action of a parameterized quantum circuit (PQC); that is, a circuit with a fixed form (a carefully chosen ansatz) but a dependence on a set of parameters $\boldsymbol{\theta}$.
For Hamiltonians $\mathcal{H}$ which can be expressed as sums over polynomially-many products of Pauli operators, the state $\ket{\Psi(\boldsymbol{\theta})}$ can be repeatedly prepared and measured $O(1/\epsilon^2)$ times to efficiently estimate the expectation value $\mathbb{E}_{\Psi,\mathcal{H}}(\boldsymbol{\theta}) = \sandwich{\Psi(\boldsymbol{\theta})}{\mathcal{H}}{\Psi(\boldsymbol{\theta})}$ to some desired precision $\epsilon$.
The variational principle gives the inequality 
\begin{eqnarray}
    \sandwich{\Psi(\boldsymbol{\theta})}{\mathcal{H}}{\Psi(\boldsymbol{\theta})} &\geq& E_0^{(\mathcal{H})},\label{eq:varprinc}
\end{eqnarray}
where $E_0^{(\mathcal{H})}$ is the ground-state energy of the Hamiltonian $\mathcal{H}$, and 
Eq. \eqref{eq:varprinc} becomes an equality only when $\ket{\Psi(\boldsymbol{\theta})}$ is a ground-state of the Hamiltonian $\mathcal{H}$.
Thus, if the ansatz $U(\boldsymbol{\theta})$ is sufficiently expressive, minimising the expectation value $\mathbb{E}_{\Psi,\mathcal{H}}(\boldsymbol{\theta})$ over the parameters $\boldsymbol{\theta}$ will yield a good approximation to the ground-state $\ket{\Psi_0^{(\mathcal{H})}}$.
The minimization can be performed by inserting the quantum processor into a feedback loop that incorporates a classical optimization algorithm; the original proposal \citep{peruzzo2014variational} used the Nelder-Mead simplex method \citep{nelder1965simplex}, but more recent work has used other optimization methods (see the reviews \citep{cao2019quantum,tilly2021variational} for examples).
Fig. \ref{fig:vqe_diagram} shows a schematic of the VQE algorithm.

\subsubsection{Quantum Approximate Optimization Algorithm}

The Quantum Approximate Optimization Algorithm (QAOA) is an algorithm for combinatorial optimization first proposed by \cite{farhi2014quantum} and which quickly inspired a number of others works developing the idea further \citep{farhi2014quantum2,farhi2016quantum,yang2017optimizing,jiang2017near,wang2018quantum,hadfield2019quantum}.
While some forms of QAOA may be considered to be purely quantum algorithms \citep{farhi2014quantum}, it is typically hybrid in form.
In fact, QAOA in its most common form can be considered to be an example of a VQE algorithm applied to combinatorial optimization, with an ansatz inspired by QA (see subsection \ref{ssec:annealing}).
The ansatz usually takes the form 
\begin{eqnarray}
    &\ket{\Psi(\boldsymbol{\alpha}, \boldsymbol{\beta})}_n = \prod_{j=1}^p\left[ \exp\left(-i\alpha_j H_D\right)\times \right. \nonumber\\
    &\left. \exp\left(-i\beta_j H_P\right)\right]\ket{\mathrm{gs}(H_D)}_n & \label{eq:qaoa_ansatz}
\end{eqnarray}
where $p$ is the number of QAOA layers and $\boldsymbol{\alpha}=(\alpha_1,\dots,\alpha_p)$, $\boldsymbol{\beta}=(\beta_1,\dots,\beta_p)$ are $2p$ variational parameters.
The classical `problem Hamiltonian' $H_P$ is a Hamiltonian into which, as in QA, a combinatorial optimization cost function has been mapped such that its eigenstates corresponds to candidate solutions and its eigenvalues correspond to their costs.
The `driver' or `mixing' Hamiltonian $H_D$ is a Hamiltonian which induces transitions between basis states.
As in QA, this is typically a simple single-body transverse field Hamiltonian $H_D=-\sum_{k=1}^n X_k$ applied to the qubits, although this has been generalized \citep{hadfield2019quantum} to include a much broader class of mixing Hamiltonians, in a similar algorithm known as the Quantum Alternating Operator Ansatz, sharing an acronym with the original algorithm.
In a typical QAOA implementation the initial state $\ket{\mathrm{gs}(H_D)}$ is the ground-state of the mixer Hamiltonian $H_D$.

The ansatz in Eq. \eqref{eq:qaoa_ansatz} is inspired by the QA algorithm in the sense that the parameters $\boldsymbol{\alpha}, \boldsymbol{\beta}$ can be chosen such that the unitary that prepares the ansatz can be expressed as a Trotterization of a QA protocol; that is
\begin{eqnarray}
    \lim_{p\rightarrow\infty}\left[\prod_{j=1}^p \exp\Bigg(-i \left(1 - s\left(\frac{jT}{p}\right)\right) H_D\Bigg) \right. \times & & \nonumber\\
    \left. \exp \Bigg(-i s\left(\frac{jT}{p}\right) H_P\Bigg) \right] = & & \nonumber\\
    \mathcal{T}\exp\left[\intop_{t=0}^{t=T}\mathrm{d}t \Big( (1-s(t)) H_D + s(t) H_P \Big) \right]. & &
\end{eqnarray}

In practice, running a QAOA algorithm with a large number $p$ of layers in order to approach the QA limit requires a large circuit depth that quickly becomes impractical, particularly in the NISQ era.
Instead, QAOA algorithms are typically considered at a small number of layers, even sometimes $p=1$ (e.g in \citep{wang2020xy}).
For a small number of layers, there are a manageable number of parameters $\boldsymbol{\alpha}, \boldsymbol{\beta}$ which can be optimized in a classical variational loop as with other variational algorithms, as described in section \ref{sssec:vqe}.
Other strategies for optimizing the parameters have also been considered, such as posing the optimization as a learning and task and employing reinforcement learning \citep{wauters2020reinforcement} to find good parameters that can even be transferred between problem instances of different sizes.

\subsubsection{Difficulties with variational algorithms}

The performance of a variational algorithm, including QAOA, will depend strictly on the choice of ansatz for the PQC $U(\boldsymbol{\theta})$.
Most critically, if the ansatz is too simple, the set of accessible states $\{\ket{\Psi(\boldsymbol{\theta}}):\forall \boldsymbol{\theta}\}$ may exclude a large portion of the physically-relevant state-space, meaning that it is not possible to achieve a good approximation to the target state $\ket{\Psi_0^{(\mathcal{H})}}$.
This can be avoided by increasing the number of independent parameters $\boldsymbol{\theta}$ in the ansatz, but this comes at the cost of increased circuit depth, which for NISQ devices will lead to increased noise, again limiting the quality of achievable solutions.

In principle, matching the form of the ansatz to the problem under consideration can also improve VQE performance; for example, the unitary coupled cluster ansatz is known to be particular suitable for problems in quantum chemistry \citep{peruzzo2014variational,yung2014transistor,romero2018strategies}.
However, such domain-specific ansatze often require large circuit depths to implement in terms of gates native to quantum processors.
An alternative approach is to employ ``hardware-efficient ansatze", which are designed to be easy to implement on real hardware \citep{peruzzo2014variational,kandala2017hardware}.
Unfortunately, the hardware-efficient approach has been observed to lead to ``barren plateaus", regions of the optimization landscape with vanishing gradients, making the classical optimization much harder to perform.
A useful discussion of ansatz design is available in \cite{mcardle2020qantum}. 

An understanding is emerging that the origin of the barren plateaus relates to the expressivity of the ansatz, \citep{holmes2021connecting}. 
In particular, if an ansatz is too expressive then the broadness of the search of the space leads to barren plateaus since the majority of the solution space will consist of low quality solutions which are relatively unaffected by small changes. 
This leads to tension between having an ansatz which is expressive enough to accurately describe a desired solution, but not so expressive that it is no longer trainable. 
It has also been found that there is a distinct mechanism by which noise in the circuit can lead to barren plateaus \citep{wang2021noise}. 
The fact that barren plateaus cause gradients to vanish begs the question whether they will be problematic if optimization of parameters is performed with a gradient free method, unfortunately, recent work suggests that they are still an impediment \citep{arrasmith2021effect}. 
While barren plateaus are a significant obstacle to variational algorithms, the situation is far from hopeless, and there are strategies which have shown some success. 
For example, it has been demonstrated in \cite{Volkoff2021correlate} that reducing the size of the search space by correlating circuit parameters can lead to large gradients (therefore destroying the plateaus). 
While many error mitigation strategies do not improve the trainability of noisy variational algorithms, some can \citep{wang2021mitigate}. 
Furthermore, progress may be made by choosing ansatze or starting conditions which are inspired by the problem being solved. 
For example, it is known that QAOA can describe digitized QA, and indeed optimal QAOA protocols have been found to be very similar (but not identical) to digitized annealing \citep{brady2021optimal,brady2021behavior} and sophisticated mathematical tools have been developed to understand when an ansatze will and will not be trainable due to barren plateaus \citep{larocca2021diagnosing}.

A related, but distinct question is how to encode information which is not naturally binary into a quantum computer. 
In particular this raises the question of binary encodings, where bitstrings are used directly to represent configurations, or encoding individual variables using unary (qubits required scale linearly with the number of configurations) encodings. 
In the case of annealing, the inability to efficiently engineer higher order terms means that it can be shown mathematically that a kind of unary encoding (domain-wall) is the most efficient method in terms of qubit usage for completely general interactions \cite{berwald2021understanding}. 
However, for gate model approaches, the tradeoff is far more complicated and encoding strategies remains an area of active research \cite{sawaya2020encoding,fuchs2021encoding,campbell2022qaoa,plewa2021variational} with many open questions. 
For example, to the best of our knowledge only Plewa et.~al.~\cite{plewa2021variational} have performed simulations to test the domain-wall encoding in a gate model setting.

\subsection{Quantum Amplitude Estimation\label{subsec:QAE}}

Quantum Amplitude Estimation (QAE) algorithms are collection of quantum methods to solve the following problem.
Given black-box access to some unitary quantum algorithm $A$ on $n$ qubits a unitary oracle $O$ that partitions the computational basis states into a set of `good' states $G$ and `bad' states $B$, 
\begin{eqnarray}
    O\ket{j} &=& \left\{
    \begin{array}{lr}
        \ket{j}, & \mathrm{for\ }j \in G \\
        -\ket{j}, & \mathrm{for\ }j \notin G,
    \end{array}\right.
\end{eqnarray}
find the probability (referred to as the \textit{amplitude} in this context) $a$ that a computational basis measurement on the output of the algorithm $\ket{\psi}=A\ket{\psi_0}$, where $\ket{\psi_0}$ is an easily-prepared reference state (typically the all-zero state $\ket{0}_n$), produces a `good' state.

Given no other information about the structure of the algorithm $A$ or the `good'/`bad' partition, the only option to solve this problem classically (aside from the unavoidable quantum run of the algorithm $A$ and the call to the oracle $O$ to check the state) is to simply run the algorithm $A$, measure the output state, check if it is `good', and repeat to build up statistics.
To achieve an estimate $\tilde{a}$ of the amplitude $a$ to within an error $\varepsilon$, this approach requires a number of calls $N^{(\mathrm{sampling})}_\mathrm{calls}$ to the algorithm $A$ that scales as
\begin{eqnarray}
    N^{(\mathrm{sampling})}_\mathrm{calls} &\in& O\left( \frac{1}{\varepsilon^2} \right)\label{eq:classical_ae}.
\end{eqnarray}

The first QAE algorithm was proposed in \cite{brassard2002quantum}, and achieves a quadratic speed-up over classical sampling; that is, the required number of calls $N^{(\mathrm{QAE})}_\mathrm{calls}$ to the algorithm $A$ to achieve an estimate $\tilde{a}$ of the amplitude $a$ to within an error $\varepsilon$ scales as
\begin{eqnarray}
    N^{(\mathrm{QAE})}_\mathrm{calls} &\in& O\left( \frac{1}{\varepsilon} \right).\label{eq:qae_speedup}
\end{eqnarray}

It was later shown that practical implementations of a QAE algorithm would have the potential to speed-up a wide range of important applications; in particular, it was shown by \cite{montanaro2015quantum} that QAE can be used to achieve a quadratic (up to logarithmic factors) speed-up for Monte Carlo algorithms by enhancing the sampling convergence rate.
Given the importance of Monte Carlo algorithms in a broad range of areas, from chemistry \citep{hammond1994monte} to statistical physics \citep{krauth2006statistical} to computational finance \citep{glasserman2004monte}, practical implementations of QAE algorithms would have a significant impact.

The QAE algorithm of \citep{brassard2002quantum} is a purely quantum algorithm, relying on the phase-estimation procedure similar to that shown in Fig. \ref{fig:phaseestcirc}, and involves many complicated controlled unitary operations that may be difficult to implement.
After the work in \citep{montanaro2015quantum} showed potential uses of QAE, many new forms of QAE were developed \citep{aaronson2020quantum,suzuki2020amplitude,nakaji2020faster,grinko2021iterative} that replaced the phase-estimation procedure with classical post-processing, but the core workings of the algorithm remained the same.
All forms of QAE involve constructing Grover-like iteration operators \citep{grover1996fast}
\begin{eqnarray}
    Q &=& (A(\mathbb{1} - 2\ket{0}_n\bra{0}_n)A^{\dagger})O,
\end{eqnarray}
similar to Eq. \eqref{eq:grover}.
The algorithms work by repeating the Grover iteration operator different numbers of times to amplify the amplitude $a$ by different amounts, and detecting which numbers of iterations produce a large probability of measuring `good' states.
The original algorithm of \cite{brassard2002quantum} uses phase-estimation to do this, while more recent QAE algorithms make use classical post-processing instead.

An illustrative example of a hybrid QAE algorithm is the Maximum-Likelihood QAE (MLQAE), proposed by Suzuki et al. \cite{suzuki2020amplitude}.
In MLQAE, Grover-type circuits are run at various different numbers $m$ of iterations to produce the states
\begin{eqnarray}
    \ket{\psi^{(m)}} &\equiv& (Q^m)A\ket{\psi_0} \\
    & = & \sin((2m+1)\theta_a)\frac{\Pi_G\ket{\psi_0}}{||\Pi_G\ket{\psi_0}||} + \nonumber\\ 
    & &\cos((2m+1)\theta_a)\frac{\Pi_B\ket{\psi_0}}{||\Pi_B\ket{\psi_0}||}
\end{eqnarray}
where the angle $\theta_a$ is defined by $\sin^2\theta_a \equiv a$ and $\Pi_G$ ($\Pi_B$) is a projection operator on to the `good' (`bad') subspace of the computational basis.
Then, these output states are measured, and it is recorded whether the outcome is `good' or `bad'.
This process is repeated for some number $N_\mathrm{shot}$ of shots, and a measurement record $\boldsymbol{h}=(h_m)_m$ is constructed, where $h_m$ is the number of `good' states produced by $N_\mathrm{shot}$ shots of the circuit with $m$ Grover iterations.

\begin{figure}
\subfigure[]{\includegraphics[width=0.99\columnwidth]{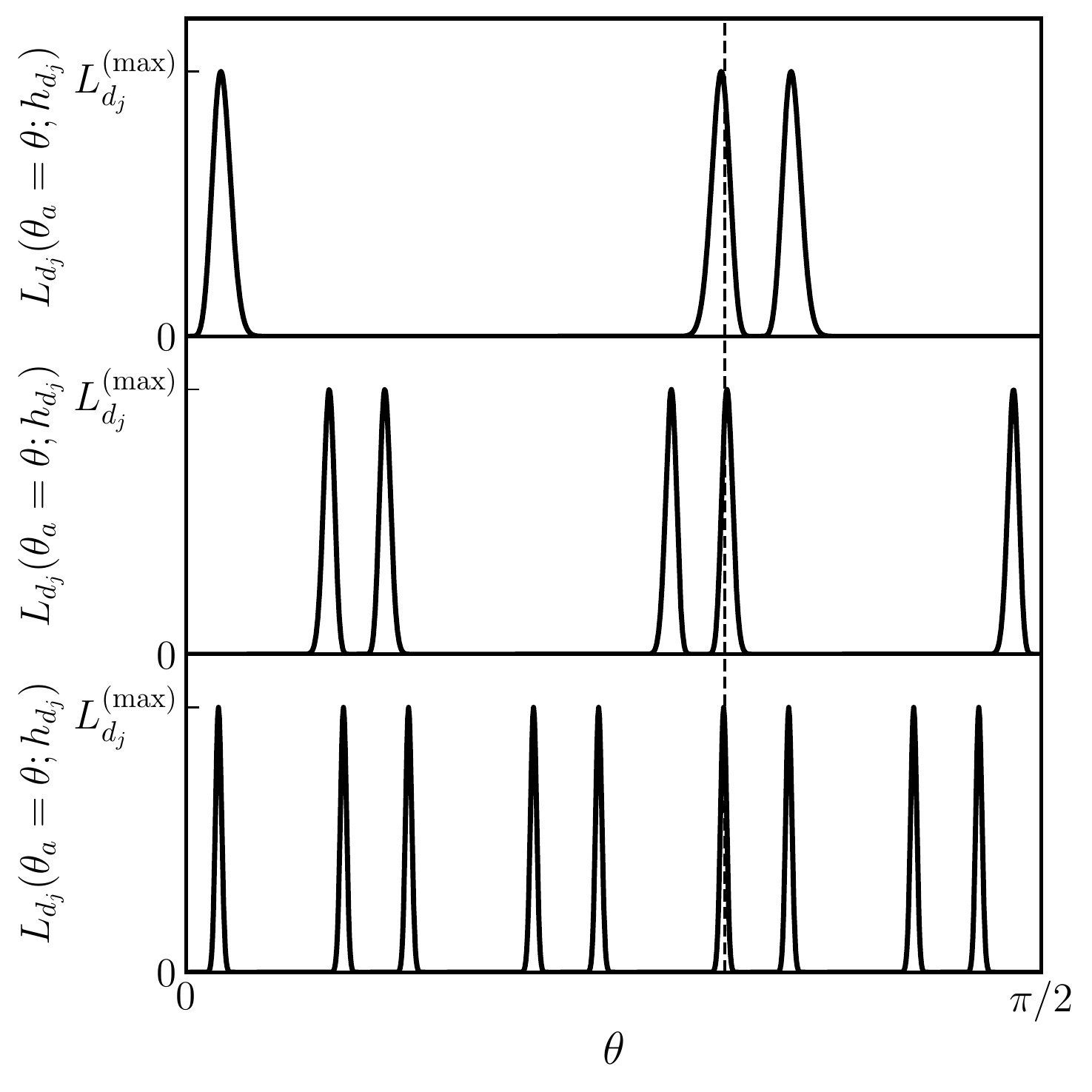}\label{fig:mlqae_likelihooda}}
\subfigure[]{\includegraphics[width=0.99\columnwidth]{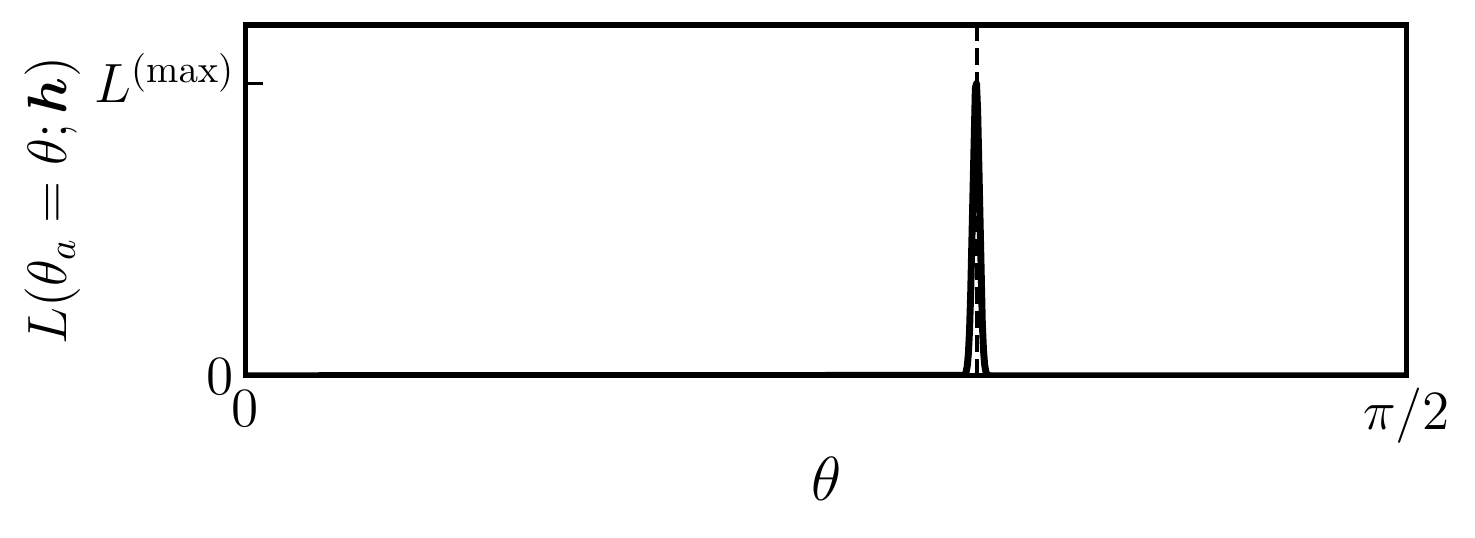}\label{fig:mlqae_likelihoodb}}
\caption{\label{fig:mlqae_likelihood} 
(a) Illustrations of the likelihood $L_m(\theta_{\tilde{a}};h_m)$ of a Grover-like QAE circuit with $m$ Grover iterations ($m$ larger for lower plots) to produce $h_m$ `good states' from $N_\mathrm{shot}$ runs, with the angle $\theta_{\tilde{a}}$.
The actual angle $\theta_a$ that generated the outcomes is indicated by the dashed line.
(b) The overall likelihood $L_m(\theta_{\tilde{a}};\boldsymbol{h})$ produced by multiplying together the likelihoods for each $m$.
}
\end{figure}

A likelihood function $L(\theta_{\tilde{a}};\boldsymbol{h})$ for the measurement record is then built,
\begin{eqnarray}
    L(\theta_{\tilde{a}};\boldsymbol{h}) &=& \prod_m L_m(\theta_{\tilde{a}};h_m)\\
    & = &\prod_m \left(\sin^2((2m+1)\theta_{\tilde{a}})\right)^{h_m}\times\nonumber\\
    & &\left(\cos^2((2m+1)\theta_{\tilde{a}})\right)^{(N_\mathrm{shot}-h_m)}.
\end{eqnarray}
By classically maximising this likelihood function (or its logarithm), the angle $\theta_a$, most likely to produce the measurement record can be determined, and hence an estimate $\tilde{a}$ for amplitude $a$ can be found.
This is illustrated in Fig. \ref{fig:mlqae_likelihood}.
Fig. \ref{fig:mlqae_likelihooda} shows an illustration of the likelihood $L_m(\theta_{\tilde{a}};h_m)$ of a Grover-like QAE circuit with $m$ Grover iterations ($m$ larger for lower plots) to produce $h_m$ `good states' from $N_\mathrm{shot}$ runs, with the angle $\theta_{\tilde{a}}$.
It can be seen that smaller numbers $m$ of iterations produce a small number of broad peaks (a small number of rough regions where the true $\theta_a$ might be), while larger $m$ values produce many sharp peaks (which can help identify an accurate estimate if the rough region is known).
The overall likelihood function, illustrated in Fig. \ref{fig:mlqae_likelihoodb}, can be maximized to find an accurate estimate of the angle $\theta_a$ (and hence the amplitude $a$).
It can be shown \citep{suzuki2020amplitude} that for some sequences of choices of $m$, MLQAE can achieve the same quadratic speedup (Eq. \eqref{eq:qae_speedup} compared to the scaling of classical sampling (Eq. \eqref{eq:classical_ae}).

Various modifications to MLQAE have been developed, including dealing with limited circuits depths and with noisy circuits \citep{giurgica-tiron2020low,brown2020quantum,uno2021modified}, but all versions offer a clear illustration of the power of (non-trivial) classical and quantum processing working together to produce speedups over what is possible by classical processing alone.

\subsection{Quantum subroutines within quantum chemistry and materials simulations}
A promising direction of research for early quantum algorithms is the development of quantum methods for solving problems in quantum chemistry.
Systems for which quantum mechanics must be taken into account to calculate important properties tend in the general case to be hard to treat with purely classical techniques; this is due to the exponential growth of the size of the Hilbert space compared to the system size.
Directly exploiting the quantum nature of quantum computers could lead to very natural techniques for these problems.

One approach that is the subject of much recent research \citep{bauer2020quantum, cerezo2021variational} is to apply the hybrid VQE algorithm described in subsubsection \ref{sssec:vqe}.
By choosing the Hamiltonian $\hat{H}$ used in the VQE loop to be the Hamiltonian of the chemical system under study (as well as suitable VQE ansatz), the VQE procedure can be used to find the ground state (or other eigenstates) of the Hamiltonian, from which other important properties can be calculated (e.g the Green's function \citep{bauer2016hybrid}, or various chemical properties.

While the direct VQE approach may be suitable for problems that are small enough to fit on near-term quantum devices (but are nonetheless large enough to be difficult for classical methods), for larger problems (such as for correlated materials simulations) a more involved approach must be taken.
One proposal from Ma et. al \citep{ma2020quantum} considers separating a full system into a small active space, with an effective quantum Hamiltonian, and its environment, which can be treated less accurately via density functional theory (DFT).
The small active space can then be treated with VQE to learn about the electronic properties of the whole system.

Many purely classical algorithms for quantum chemistry and materials rely on quantum Monte Carlo \cite{gull2011continuous,austin2012quantum}, a family of classical algorithms which can, to some extent, emulate quantum systems in thermal equilibrium. 
Quantum Monte Carlo is limited for fermionic systems, however, by what is known as the sign problem, where accurate sampling can become computationally infeasible below certain temperatures \cite{loh1990sign}.
These algorithms have the advantage of providing ready-made classical wrappers to add in quantum components, and are therefore natural candidates for early hybrid quantum-classical algorithms, such as the example presented in \cite{bauer2016hybrid}. 
While a full review of hybrid quantum-classical materials and chemistry algorithms, including methods which are based on time evolution rather than ground state preparation is beyond the scope of our current work, Bauer et al. review the algorithmic opportunities in \cite{bauer2020quantum}.

To give a flavour of how these algorithms work, we focus on work by Bauer et al. \cite{bauer2016hybrid} which addresses the significant computational challenge to simulate the properties of strongly-correlated materials where the independent-electron approximation intrinsic to the standard DFT breaks down.
This work explores the dynamical mean field theory (DMFT) method (see \citep{georges1996dynamical} for a thorough review), which can be applied to materials expressible as lattice models.
The DMFT approach which recasts the material as a local site (known as an `impurity') interacting with the rest of the system expressed as a non-interacting bath.
The total Hamiltonian is written as 
\begin{eqnarray}
    H_\mathrm{DMFT} &=& H_\mathrm{imp} + H_\mathrm{bath} + H_\mathrm{mix}
\end{eqnarray}
where 
\begin{eqnarray}
    H_\mathrm{imp} &=& \sum_{a,b=1}^N t_{ab} f^\dagger_a f_b + \nonumber \\
    && \sum_{a,b,c,d=1}^N U_{abcd} f^\dagger_a f^\dagger_b f_c f_d
\end{eqnarray}
is the impurity Hamiltonian that will be treated carefully, where $f^\dagger$ ($f$) operators create (annihilate) a fermion in one of $N$ local orbitals and the hopping and interaction integrals $t_{ab}$ and $U_{abcd}$ are determined from the underlying material.
The non-interacting bath Hamiltonian has the form
\begin{eqnarray}
    H_\mathrm{bath} &=& \sum_j \epsilon_j b_j^\dagger b_j, \\
\end{eqnarray}
where the $b^\dagger$ ($b$) operators create (annihilate) a fermion in one of the bath modes (considered to be infinitely many, but can be truncated to a finite number of modes if necessary).
The remaining term,
\begin{eqnarray}
    H_\mathrm{mix} &=& \sum_{a=1}^N\sum_j \left(V_{aj} f^\dagger_a b_j + V_{aj}^* b^\dagger_j f_a\right), \\
\end{eqnarray}
describes the interaction between the impurity and the bath.
While at this level of description the bath mode energies $\epsilon_j$ and interactions $V_{aj}$ are free parameters, the core of the DMFT algorithm is to iteratively update these parameters in a self-consistent way until the Green's function of the impurity model and the relevant local terms of the lattice Green's function converge to match.
This DMFT scheme is used within an additional outer loop that uses the information it provides about the correlated electrons to iteratively improve a DFT solution.

\begin{figure}[ht]
\subfigure[]{\includegraphics[width=0.99\columnwidth]{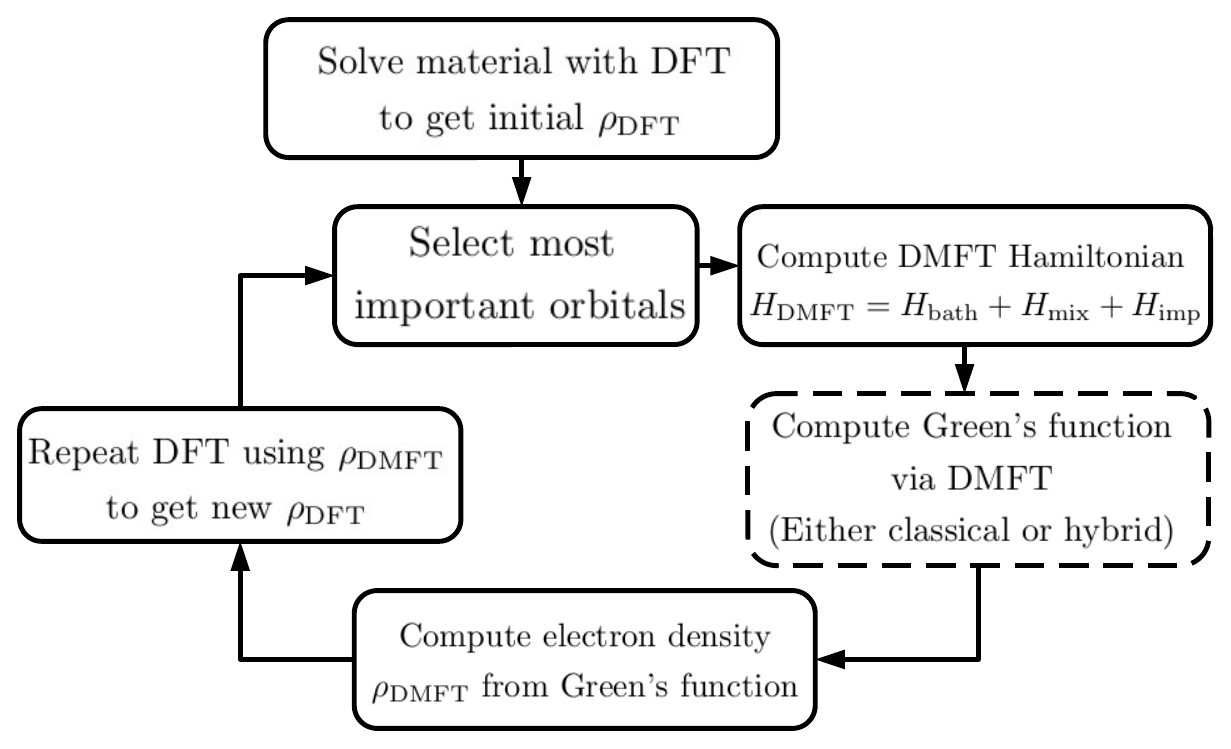}\label{fig:dmftdft_outer}}
\subfigure[]{\includegraphics[width=0.99\columnwidth]{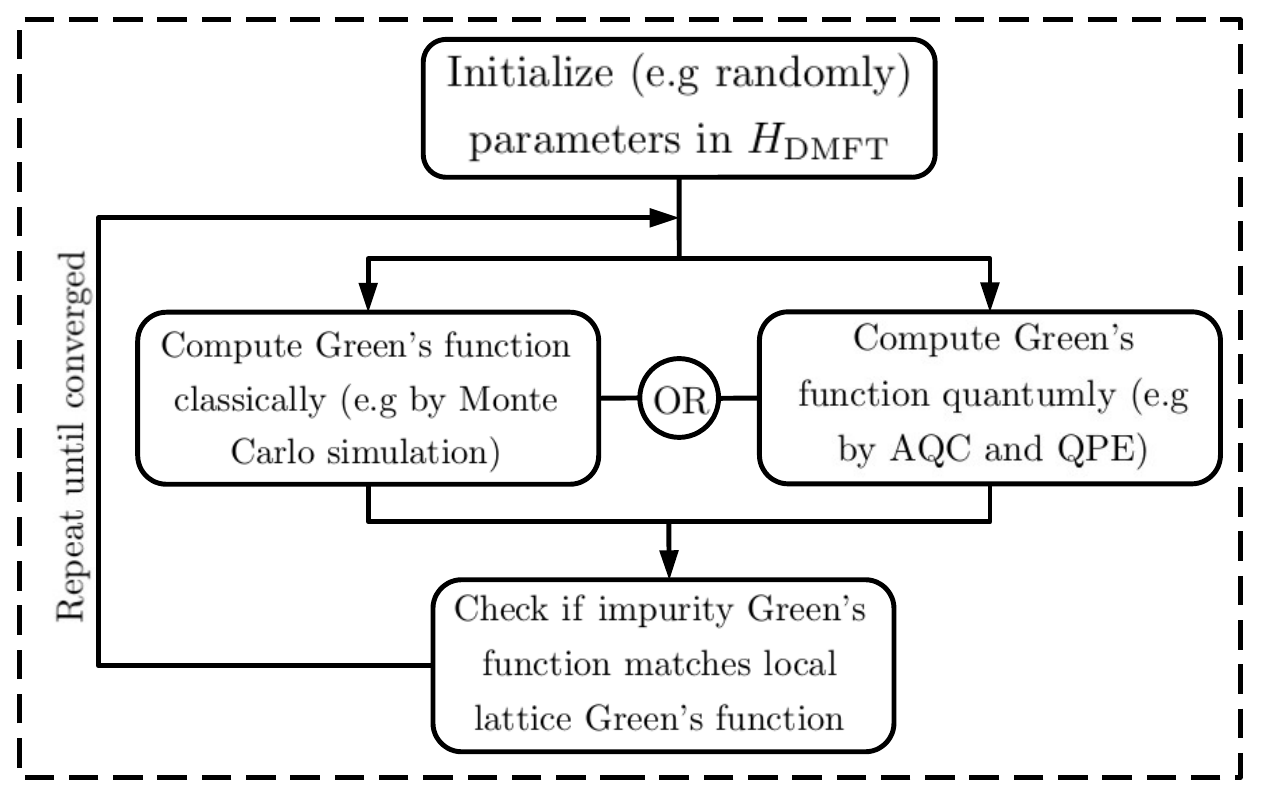}\label{fig:dmft}}
\caption{\label{fig:dmftdft_all} 
Schematics showing the high-level structure of the hybrid algorithm for correlated materials proposed in \citep{bauer2016hybrid}.
(a) The outer loop in which a DFT solution is iteratively improved.
The step in which the Green's function for the impurity model is calculated is shown with a dashed border to indicate that it is \textit{this} piece of the algorithm that can be swapped out for a quantum subroutine.
(b) The expanded version of the impurity model Green's function calculation from Fig. \ref{fig:dmftdft_outer}, indicating that it can be achieved either purely classically or on a hybrid quantum-classical way.
}
\end{figure}

The structure of the outer loop is shown in Fig. \ref{fig:dmftdft_outer}.
The step in which the impurity Green's function is calculated is shown with a dashed border to indicate that it is \textit{this} piece of the algorithm that can be swapped out for a hybrid quantum-classical subroutine.
This step is expanded in Fig. \ref{fig:dmft}.
Classically, there are various ways to calculate the Green's function, such as by a quantum Monte Carlo algorithm as discussed previously in this section.
However, to overcome the inherent computational challenges of classical approaches, Bauer et al. \citep{bauer2016hybrid} propose using a combination of adiabatic state preparation (using AQC to `compute' approximations to Hamiltonian eigenstates) and quantum phase estimation \citep{kitaev1995quantum} to prepare the ground state, from which the Green's function can be measured.

\section{Hardware considerations \label{sec:hardware}}

\subsection{Larger computations on smaller hardware - Entanglement forging}
Since near-term quantum processors will have only a limited number of qubits available, it is worthwhile to consider how classical processing can be used to apply hybrid algorithms to problems that may not otherwise fit on small devices. 
An illustrative example of such a method is the recent \textit{entanglement forging}, due to \cite{eddins2022doubling}.
In that work, Eddins et al. show how a quantum system mapped to $2N$ qubits can be studied on a device with only $N$ qubits, by breaking the $2N$-qubit system into two $N$-qubit partitions and expressing the density operator $\rho=\ketbra{\Psi}{\Psi}$ of the system in terms of the Schmidt decomposition
\begin{eqnarray}
    \ket{\Psi} & = & (U \otimes V)\sum_{k=1}^{2^N -1}\lambda_k \ketbra{b_k}{b_k}^{\otimes 2}
\end{eqnarray}
\sloppy where $\ket{b_k}$ are computational basis states for an $N$-qubit, $U,V$ are unitary operators local to each half of the bipartition. The non-negative,real-value Schmidt coefficients $\lambda_k$ are treated as variational parameters for a VQE circuit (see subsection \ref{sssec:vqe}).
In this representation, expectation values $\sandwich{\Psi}{O}{\Psi}$ of $2N$-qubit operators of the form $O=O_1\otimes O_2$ can be decomposed into terms of the form $\sandwich{\phi_1}{UO_1U^\dagger}{\phi_1}\sandwich{\phi_2}{VO_2V^\dagger}{\phi_2}$, where $\ket{\phi}$ can be single basis states $\ket{b_k}$ or uniform superpositions of two basis states. 
These expectation values can be computed on $N$ qubits, and are then sampled in a weighted way according to the coefficients $\lambda_k\lambda_j$.

While this will not scale well in the general case, systems that are only weakly-entangled across the bipartition will have only a small number of dominant Schmidt coefficients $\lambda_k$; thus the expectation value $\sandwich{\Psi}{O}{\Psi}$ can be well-approximated with a manageable number of samples.
In \cite{eddins2022doubling}, entanglement forging (along with some heuristic enchancements) is used to find ground-state energies of various geometries of the water molecule with high-accuracy, via a ten-qubit mapping, usimg only 5 superconducting qubits on real quantum hardware. 

\subsection{Additional controls}

In the case of QA and other non-universal analog processors, we see that sometimes hybrid techniques required additional controls. 
The most prototypical example here is reverse annealing as discussed in subsubsection \ref{subsubsec:hybrid_anneal}, which allows for hybrid techniques. 
These controls are important because they allow a much richer variety of techniques which have demonstrated real gains in experiments. 
In particular, in this example the controls were actually already available within the underlying device (the protocol described in \cite{lanting2014entanglement} is essentially reverse annealing combined with some other techniques), but had not been made available to external users. It is therefore highly important to think about how existing experimental techniques could be used in computation for different purposes than they were originally intended.

\subsection{Co-locating the quantum and classical computing power}

When discussing computations at the abstract level we have so far, it is easy to forget that communication between classical and quantum systems is not free.
While the currently dominant model of cloud access to quantum computers is very useful for proof-of-concept experiments, the time required to send signals over the internet between quantum and classical components could prove to be a major barrier. 
There are already efforts to co-locate quantum computing resources with powerful classical high performance computing, for example through the J{\"u}lich supercomputing centre's JUNIQ program \citep{juniq2019launch}. 

Some computation may require even faster communication between quantum and classical components; for qubit technology such as superconducting qubits which operate on chip substrates at ultra-low temperatures, this means that classical computation at cryogenic temperatures (although possibly not as cold as the quantum computer) could be beneficial. 
Inspired by the quantum computing use cases, there have been numerous studies in this direction \cite{homulle2018cryogenic,patra2018cryo,weinreb2007design,homulle2018performance,conway-lamb2016fpga}. 
As we discuss in section \ref{sec:specialized_accelerators}, there are also many types of classical processors, each with their strengths and weaknesses; any of these could potentially be useful candidates to co-locate with quantum computers.

\section{The big picture: specialized accelerators \label{sec:specialized_accelerators}}
Throughout this work, we have offered many examples showing how classical and quantum processors can potentially be used together to perform computational tasks much more efficiently than what is possible with a classical processor alone.
While a quantum processor is certainly quite different, both practically and theoretically, from any kind of classical processor, different types of processors working together is already a common structure within purely classical computing; there are many classical examples of specialized co-processors, to which specific kinds of processing are offloaded, and this is sometimes known as \textit{heterogeneous} or \textit{heterotic}.
A thorough discussion of heterotic computing is available in \cite{stepney2012framework}.
In this section, we will briefly describe some examples of classical heterotic computing and offer some intuition for how quantum processors can naturally fit into this framework.

\paragraph*{Graphics Processing Unit -}
The clearest example of classical heterogeneous computing is that of a computer's central processing unit (CPU) working together with graphics processing unit (GPU) \citep{mcclanahan2011history}.
A typical GPU consists of, among other things, a much larger number of cores than a typical CPU, and these cores are particularly suited to the kinds of three-dimensional vector-matrix operations needed for graphics processing, such as translations and rotations in space.
Typically, the same calculation needs to happen many different times on independent inputs; for example, every point on a three-dimensional object needs to move in the same way, with only the starting conditions different for each point.
The large number of small cores on a GPU allows this to happen to happen at a greatly accelerated pace \citep{mcclanahan2011history} compared to the relatively small number of (more general purpose) cores on a CPU can manage.
Typically, the computation is arranged such that the CPU hands off these specific tasks to the GPU when needed.

More recently, GPU technology has been used for applications beyond graphics processing but which also require the same large-scale parallelization of small numerical operations, such as machine learning calculations (e.g via TensorFlow \citep{abadi2016tensorflow}).

\paragraph*{Application-specific integrated circuit -}
An Application-specific integrated circuit (ASIC) \citep{smith1997application} can be thought of as a processor designed and manufactured to carry out a specific computational task much more quickly and with much more energy efficiency than can be achieved by a programming a general-purpose processor such as a CPU.
An example of an ASIC with these properties is Intel's Quick Sync video technology; these circuits are used for rapidly converting video between different formats, and are integrated with Intel CPUs to handle these specific workloads \citep{angelini2011intel}.
Similar hardware, known as Field-programmable Gate Arrays (FPGAs) \citep{trimberger2015three}, can be reprogrammed at the hardware level after being deployed.

\paragraph{Neuromorphic processors -}
Many modern classical algorithms are based mimicking the behaviour of the human brain through the use of software-defined neural networks.
While this approach is proving successful in many areas, standard processors are proving to be limited by particular bottlenecks, such as the communication between the core and memory.
To avoid these limitations, the emerging development of neuromorphic processors \citep{james2017historical} aim to emulate the behaviour of the human brain directly at the hardware level, leading to hardware that can run artificial intelligence applications much more efficiently than standard processors.

\paragraph{Quantum computers as specialized co-processors -}

The specialized accelerators described here all show how hardware designed and optimized for a specific type of computational work can operate in tandem with a standard processor, such as a CPU, to enhance overall computational efficiency.
Hybrid quantum-classical algorithms, including the ones described in this work, fit naturally into this framework; a quantum processing unit (QPU) can be called by a general purpose classical CPU to accelerate specific computational subroutines, such as the phase estimation component of Shor's algorithm (see subsection \ref{ssec:shor}) or the ansatz preparation component of a VQE algorithm (see subsubsection \ref{sssec:vqe}).
From this perspective, hybrid quantum-classical algorithms are a continuation of a traditional pattern in the development of computer technology. 
While the operating principles of a quantum computer are quite different from their classical counterparts, the way they fit into larger architectures (and the surrounding concerns) are not fundamentally something which has not been encountered before.

\section{The future of hybrid algorithms \label{sec:future}}

The development path of hybrid algorithms is not universal and does not always involve a non-hybrid precursor; while hybrid forms of QAE grew out of the original pure form of QAE (and indirectly out of Grover's search, a pure quantum algorithm), many others came about in different and varied ways. 
Shor's algorithm for example emerged from the fact that a quantum algorithm fulfilled a niche application which allowed for a very powerful computational tool.
This is similar to many hybrid materials and chemistry algorithms, which have developed out of classical algorithms by replacing classical methods such quantum Monte Carlo with quantum computing calls. 
In the case of variational algorithms, as well as hybrid forms of quantum annealing, the capabilities of the hardware has driven the development of algorithms. 
A key lesson to draw from these stories is that there is not one way to develop hybrid algorithms: they can emerge in a multitude of different ways.

An immediate question which arises from this discussion is whether hybrid algorithms are just a phase in the evolution of quantum computing, and when (and if) large, fully fault-tolerant quantum computers come into existence, will hybrid algorithms still be used? 
If we look at the evolution of classical computing technologies, we can find a hint. 
As new, advanced co-processors have been developed classically they have not displaced older paradigms, but rather been added to them to perform specialized tasks.
We do not see any reason to believe that the evolution of quantum computers will be any different. 
Quantum computers are not intrinsically faster at simple applications like adding numbers, and even if large numbers of high-quality qubits became readily available, it is doubtful that they would ever be better than classical bits by all metrics, in which case heterotic computing would be the most sensible paradigm, as it currently is in the world of classical computing. 
We find it unlikely that once quantum hardware advances to the stage of genuine usefulness 
non-hybrid algorithms will ever be deployed in a ``production'' setting, for the simple reason that it is likely that further improvements could be made by adding a hybrid component.

From our discussion so far, we have seen that hybrid algorithms are most meaningfully classified in terms not of how much classical computing power they use, but how that computation fits into the overall model. 
While there are notable algorithms which are both hybrid and non-hybrid, a pattern which has appeared both in terms of Grover's algorithm and subsequent QAE techniques, and quantum annealing and subsequent hybrid versions, is that non-hybrid algorithms often evolve into richer, more powerful, hybrid variants. 
This pattern is likely not incidental, it is often easier to first imagine algorithms in a more ``pure'' setting and then allow them to evolve. 
This suggests an interesting relationship between hybrid and non-hyrbid algorithms: the non-hybrid setting is a useful incubator for hybrid techniques which will later be implemented. 
As such, we are not suggesting that hybrid algorithms are the only class of quantum algorithms which are worthwhile to study, but that it is probably most appropriate to think of non-hybrid algorithms as the foundations of new algorithms which are yet to be developed. 

\section{Acknowledgments}

The authors thank Raouf Dridi, Viv Kendon, Susan Stepney, and Dan Browne for useful discussions and references. 
Both authors were supported by the UK Engineering and Physical Science Research Council (EPSRC)) grant number EP/T026715/1 for Collaborative Computational Project in Quantum Computing (CCP-QC).
NC was additionally supported by EPSRC grant EP/W00772X/1, and AC was supported by the EPSRC UK Quantum Technology Hub in Computing and Simulation (EP/T001062/1).

\bibliography{main}

\end{document}